\documentstyle[12pt,aps,prd,preprint]{revtex}
\begin{document}
\title{Late-Time Evolution of Charged Gravitational Collapse and Decay of
Charged Scalar Hair - II}
\author{Shahar Hod and Tsvi Piran}
\address{The Racah Institute for Physics, The
Hebrew University, Jerusalem 91904, Israel}
\date{\today}
\maketitle

\begin{abstract}
We study {\it analytically} the initial value problem for a {\it charged}
massless scalar-field on a Reissner-Nordstr\"om spacetime. Using the
technique of spectral decomposition we extend recent results on this problem.
Following the no-hair theorem we reveal the {\it dynamical} physical
mechanism by which the {\it charged} hair is radiated away. We show that
the charged perturbations decay according to an inverse power-law behaviour 
at future timelike infinity and along future null infinity. Along the future
outer horizon we find an oscillatory inverse power-law relaxation of the 
charged fields. We find that a {\it charged} black hole becomes ``bald'' 
{\it slower} than a neutral one, due to the existence of charged perturbations.
Our results are also important to the study of mass-inflation and the stability
of Cauchy horizons during a dynamical gravitational collapse of charged matter
in which a charged black-hole is formed.
\end{abstract}

\section{introduction}\label{introduction}
The late-time evolution of various fields outside a collapsing star plays an
important role in two major aspects of black-hole physics:

\begin{enumerate}
\item The {\it no-hair theorem} of Wheeler states that the {\it external}
field of a black-hole relaxes to a Kerr-Newman field characterized solely
by the black-hole's mass, charge and angular-momentum. Thus, it is of
interest to reveal the {\it dynamical} mechanism  responsible for the
relaxation of perturbations fields outside a black-hole. The mechanism by which
neutral fields are radiated away was first studied by Price \cite{Price}. The
physical mechanism by which a charged black-hole, which is formed during a 
gravitational collapse of a {\it charged} matter, dynamically sheds its charged
hair was first studied in paper I \cite{HodPir}. However, this analysis was
restricted to the weak electromagnetic interaction limit $|eQ| \ll1$. In this
paper we extend our {\it analytical} results to include {\it general} values
of the (dimensionless) quantity $eQ$.
 
\item The asymptotic late-time tails along the outer horizon of a rotating or
a charged black-hole are used as initial input for the ingoing perturbations 
which penetrates into the black-hole. These perturbations are the physical 
cause for the well-known phenomena of {\it mass-inflation} \cite{Poisson}. In this context,
one should take into account the existence of {\it charged} perturbations, 
which are expected to appear in a dynamical gravitational collapse of a 
{\it charged} star. Here we study the asymptotic behaviour of such charged
perturbations.
\end{enumerate}

The plan of the paper is as follows. In Sec. \ref{Sec2} we give a short
description of the physical system and formulate the evolution equation
considered. In Sec. \ref{Sec3} we formulate the problem in terms of
the black-hole Green's function using the technique of spectral decomposition.
In Sec. \ref{Sec4} we study the late-time evolution of charged scalar
perturbations on a Reissner-Nordstr\"om background. We find an inverse
{\it power-law} behaviour of the perturbations along the three asymptotic
regions: timelike infinity $i_+$, future null infinity $scri_+$ and along
the black-hole outer-horizon (where the power-law is multiplied by a 
{\it periodic} term). We find that the dumping exponents which describe the 
fall-off of {\it charged} perturbations are {\it smaller} compared with
the neutral dumping exponents. Thus, a black-hole which is formed from the
gravitational collapse of a {\it charged} matter becomes ``bald'' {\it slower}
than a neutral one due to the existence of charged perturbations. In Sec.
\ref{Sec5} we reduce our results to the Schwarzschild case (and 
equivalently, to a neutral field on a Reissner-Nordstr\"om background). We show
that one can obtain the asymptotic behaviour of the field along the outer
horizon \cite{Gundlach} using the technique of spectral decomposition. 
We conclude in Sec. \ref{Sec6} with a brief summary of our results and
their implications.

\section{Description of the system}\label{Sec2}

We consider the evolution of a massless {\it charged} scalar perturbations 
fields outside a charged collapsing star. The system was already described on
paper I \cite{HodPir}. Here we give only the final form of the 
equations studied.
The external gravitational field of a spherically symmetric collapsing star of 
mass $M$ and charge $Q$ is given by the metric

\begin{equation}\label{Eq1}
ds^2=\lambda^2 (-dt^2+dy^2)+r^2d\Omega ^2\  ,
\end{equation}
where the tortoise radial coordinate $y$ is defined by $dy=dr/\lambda^2$
and $\lambda^2={1-{{2M} \over r}+{{Q^2} \over {r^2}}}$.

Resolving the field into spherical harmonics $\phi
=e^{ie\Phi t}\sum\limits_{l,m} {\psi _m^l\left( {t,r} \right)Y_l^m{{\left( {
\theta ,\varphi} \right)} \mathord{\left/ {\vphantom {{\left( {  \theta ,\varphi } \right)} r}}\right.}r}}$ (where $\Phi$ is merely
a gauge constant of the electromagnetic potential $A_t$, i.e. its value at
infinity) one obtains a wave-
equation for each multipole moment

\begin{equation}\label{Eq2}
\psi _{,tt}+2ie{Q \over r}\psi _{,t}-\psi _{,yy}+V\psi =0\  ,
\end{equation}
where

\begin{equation}\label{Eq3}
V=V_{M,Q,l,e}\left( r \right)=\left( {1-{{2M} \over r}+{{Q^2} \over {r^2}}}
\right)\left[ {{{l\left( {l+1} \right)} \over {r^2}}+{{2M} \over {r^3}}-{{2Q^2}
\over {r^4}}} \right]-e^2{{Q^2} \over {r^2}}\  .
\end{equation}

The electromagnetic potential $A_t$ satisfies the relation

\begin{equation}\label{Eq4}
A_t=\Phi -{Q \over r}\  ,
\end{equation}
where $\Phi$ is a constant.

\section{Formalism}\label{Sec3}

The time-evolution of a charged scalar-
field described by Eq. (\ref{Eq2}) is given by
 
\begin{equation}\label{Eq5}
\psi (y,t)= \int \left[ G(y,x;t) \psi _t(x,0)+G_t(y,x;t) \psi (x,0)+{2ieQ 
\over {r(x)}} G(y,x;t) \psi (x,0) \right] dx\  ,
\end{equation}
for $t>0$, where the (retarded) Green's function $G(y,x;t)$ is defined as

\begin{equation}\label{Eq6}
\left [{{\partial ^2} \over {\partial t^2}} +2ie{Q \over r} {{\partial} \over
{\partial t}} -{{\partial ^2} \over {\partial y^2}} +V(r) \right ] G(y,x;t)=
\delta (t) \delta(y-x)\  . 
\end{equation}
The causality condition gives us the initial condition $G(y,x;t)=0$ for
$t \leq 0$.
In order to find $G(y,x;t)$ we use the Fourier transform 
\begin{equation}\label{Eq7}
\tilde G(y,x;w)= \int_{0^-}^{\infty} G(y,x;t) e^{iwt} dt\  .
\end{equation}
The Fourier transform is analytic in the upper half $w$-plane and it satisfies
the equation
\begin{equation}\label{Eq8}
\left ({{d^2} \over {dy^2}} +w^{2} -{{2eQw} \over r} -V \right) \tilde G(y,x;w)
= \delta(y-x)\  .
\end{equation}
$G(y,x;t)$ itself is given by the inversion formula
\begin{equation}\label{Eq9}
G(y,x;t)={1 \over {2 \pi}} \int_{- \infty +ic}^{\infty +ic} \tilde G(y,x;w)
e^{-iwt} dw\  ,
\end{equation}
where $c$ is some positive constant.

Next, we define two auxiliary functions $\tilde \psi_1(y,w)$ and 
$\tilde \psi_2(y,w)$ which are (linearly independent) solutions to the 
homogeneous equation
\begin{equation}\label{Eq10}
\left ({{d^2} \over {dy^2}} +w^{2} -{{2eQw} \over r} -V \right) \tilde 
\psi_i(y,w) =0\  ,\ \ \ \  i=1,2\  .
\end{equation}
The two basic solutions that are required in order to build the black-hole
Green's function are defined by their asymptotic behaviour:
\begin{equation}\label{Eq11}
\tilde \psi_1(y,w) \sim 
\left\{
 \begin{array}{l@{\quad,\quad}l}
e^{-i \left (w- {{eQ} \over {r_+}} \right) y} & y \to -\infty\  , \\
A_{out}(w) y^{i(2wM-eQ)} e^{iwy} + A_{in}(w) y^{-i(2wM-eQ)} e^{-iwy} & y \to \
\infty\  ,
\end{array} \right. \    
\end{equation}
and
\begin{equation}\label{Eq12}
\tilde \psi_2(y,w) \sim 
\left\{
 \begin{array}{l@{\quad,\quad}l}
B_{out}(w) e^{i \left (w- {{eQ} \over {r_+}} \right) y} +B_{in}(w)
e^{-i \left (w- {{eQ} \over {r_+}} \right) y} & y \to -\infty\  , \\
y^{i(2wM-eQ)} e^{iwy} & y \to \infty\  .
\end{array} \right.  
\end{equation}
Let the Wronskian be
\begin{equation}\label{Eq13}
W(w)=W(\tilde \psi_1, \tilde \psi_2 )= \tilde \psi_1 \tilde \psi_{2,y} - 
\tilde \psi_2 \tilde \psi_{1,y}\  ,
\end{equation}
where $W(w)$ is $y$-independent.
Thus, using the two solutions $\tilde \psi_1$ and $\tilde \psi_2$, the black-
hole Green's function can be expressed as
\begin{equation}\label{Eq14}
\tilde G(y,x;w) =- {1 \over {W(w)}} 
\left\{ \begin{array}{l@{\quad,\quad}l}
\tilde \psi_1(y,w) \tilde \psi_2(x,w) & y<x \  , \\
\tilde \psi_1(x,w) \tilde \psi_2(y,w) & y>x \  .
\end{array} \right.
\end{equation}

In order to calculate $G(y,x;t)$ using Eq. (\ref{Eq9}), one may close the 
contour of integration into the lower half of the complex frequency plane.
Then, one finds three distinct contributions to $G(y,x;t)$ \cite{Leaver} :

\begin{enumerate}
\item {\it Prompt contribution}. This comes from the integral along the large
semi-circle. It is this part, denoted $G^F$, which propagates the {\it high}-
frequency response. For large frequencies the Green's function limits to the
flat spacetime one. Thus, this term contributes to the {\it short}-time 
response (radiation which reaches the observer nearly directly from the source,
i.e. without scattering). This term can be shown to be effectively zero beyond
a certain time. Thus, it is not relevant for the late-time behaviour of the
field.

\item {\it Quasinormal modes}. This comes from the distinct singularities of
$\tilde G(y,x;w)$ in the lower half of the complex $w$-plane and is denoted
by $G^Q$. These singularities occur at frequencies for which the Wronskian
(\ref{Eq13}) vanishes. $G^Q$ is just the sum of the residues at the 
poles of $\tilde G(y,x;w)$. Since each mode has Im$w<0$ it decays 
{\it exponentially} with time. 
\item {\it Tail contribution}. The late-time tail is
associated with the existence of a branch cut (in $\tilde \psi_2$) \cite{Leaver}, usually
placed along the negative imaginary $w$-axis. This tail arises from the 
integral of $\tilde G(y,x;w)$ around the branch cut and is denoted by  $G^C$.
As will be shown, the contribution $G^C$ leads to an inverse {\it power-law}
behaviour (multiplied by a periodic term along the black-hole outer horizon) of
the field. Thus, $G^C$ dominates the late-time behaviour of the field.
\end{enumerate}

The present paper investigate the late-time asymptotic behaviour of a charged
scalar-field. Thus, the purpose of this paper is to evaluate $G^C(y,x;t)$.

\section{The late-time behaviour of a charged scalar-field}\label{Sec4}

\subsection{The large-$r$ (low-$w$) approximation}\label{Sec4A}

It is well known that the late-time behaviour of massless perturbations 
fields is determined by the backscattering from asymptotically {\it far}
regions \cite{Thorne,Price}. Thus, the late-time behaviour is dominated by the
{\it low}-frequencies contribution to the Green's function, for only low
frequencies will be backscattered by the small potential (for $r \gg M,|Q|$) in
(\ref{Eq10}). Thus, as long as the observer is situated far away from the 
black-hole and the initial data has a considerable support only far away from 
the black-hole, a {\it large}-$r$ (or equivalently, a 
{\it low}-$w$) approximation is sufficient in order to study the asymptotic {\it late-time}
behaviour of the field \cite{Andersson}.

The wave-equation (\ref{Eq10}) of the charged scalar-field (in the field of
a charged black hole) can be transformed in such a way that the Coulomb and
Newtonian $1/r$ potentials will dominate at large values of $r$. We first
introduce an auxiliary field $\xi$
\begin{equation}\label{Eq15}
\xi=\lambda \tilde \psi\  ,
\end{equation}
in terms of which the equation (\ref{Eq10}) for the charged scalar-field 
becomes
\begin{equation}\label{Eq16}
\left\{ {{d^2} \over {dr^2}} -{{\lambda_{,rr}} \over \lambda} + {1 \over
{\lambda ^{4}}} \left[ \left(w- {{eQ} \over r} \right)^2 - \lambda ^{2}
\left({{l(l+1)} \over {r^{2}}} + {{2M} \over {r^{3}}} - {{2Q^{2}} \over
{r^{4}}} \right) \right] \right\} \xi =0\  .
\end{equation}

As was explained above, one only needs a {\it large} $r$ approximation in order
to study the late-time behaviour of the charged field. Thus, we expend 
(\ref{Eq16}) as a power series in $M/r$ and $Q/r$ and obtain (neglecting terms
of order $O({w \over {r^2}})$ and smaller)
\begin{equation}\label{Eq17}
\left[ {{d^2} \over {dr^2}} +w^{2} +{{4Mw^{2}-2eQw} \over r} - 
{{l(l+1)-(eQ)^{2}} \over {r^{2}}} \right ] \xi =0\  .
\end{equation}
The terms proportional to $M/r$ and $eQ/r$ represent the Newtonian and Coulomb
potentials respective. It should be noted that this equation (with $M=0$)
represents {\it exactly} the evolution of a {\it charged} scalar-field on a
{\it charged} and {\it flat} background.

Let us now introduce a second auxiliary field $\tilde \phi$ defined by
\begin{equation}\label{Eq18}
\xi =r^{\beta +1} e^{iwr} \tilde \phi (z)\  ,
\end{equation}
where
\begin{equation}\label{Eq19}
z=-2iwr\ \ \ ;\ \ \  \beta ={{-1+ \sqrt {(2l+1)^{2} -4(eQ)^{2}}} \over 2}\  .
\end{equation}
$\tilde \phi$(z) satisfies the confluent hypergeometric equation
\begin{equation}\label{Eq20}
\left[ z {{d^{2}} \over {dz^{2}}} +(2\beta+2-z) {d \over {dz}} -(\beta+1-
2iw\alpha) \right] \tilde \phi(z)=0\  ,
\end{equation}
where
\begin{equation}\label{Eq21}
\alpha =M-{{eQ} \over {2w}}\  .
\end{equation}
Thus, the two basic solutions required in order to build the black-hole
Green's function are (for $r \gg M,|Q|$)
\begin{equation}\label{Eq22}
\tilde \psi_1 =Ar^{\beta +1} e^{iwr} M(\beta +1-2iw \alpha ,2\beta +2, 
-2iwr)\  ,
\end{equation}
and
\begin{equation}\label{Eq23}
\tilde \psi_2 =Br^{\beta +1} e^{iwr} U(\beta +1-2iw \alpha ,2\beta +2, 
-2iwr)\  ,
\end{equation}
where $A$ and $B$ are normalization constants. $M(a,b,z)$ and $U(a,b,z)$ are 
the two standard solutions to the confluent hypergeometric equation 
\cite{Abram}.
$U(a,b,z)$ is a many-valued function, i.e. there will be
a cut in $\tilde \psi_2$.

Using Eq. (\ref{Eq9}), one finds that the branch cut contribution to the 
Green's function is given by
\begin{equation}\label{Eq24}
G^C(y,x;t)={1 \over {2\pi}} \int_{0}^{-i \infty} \tilde \psi_1(x,w) \left[
{{\tilde \psi_2(y,we^{2 \pi i})} \over {W(we^{2 \pi i})}} -
{{\tilde \psi_2(y,w)} \over {W(w)}} \right] e^{-iwt} dw\  .
\end{equation}
(For simplicity we assume that the initial data has a considerable support 
only inside the observer. This, of course, does not change the {\it late}-time
behaviour, for it is a consequence of a backscattering at asymptotically
{\it far} regions).

Using the fact that $M(a,b,z)$ is a single-valued function and Eq. 13.1.10 of
\cite{Abram}, one finds
\begin{equation}\label{Eq25}
\tilde \psi_1(r,we^{2 \pi i})= \tilde \psi_1(r,w)\  ,
\end{equation}
and
\begin{equation}\label{Eq26}
\tilde \psi_2(r,we^{2 \pi i})=e^{-4i \pi \beta } \tilde \psi_2(r,w)+
(1-e^{-4i \pi \beta }) { \Gamma (-2 \beta -1) \over \Gamma (- \beta -2iw 
\alpha )} \tilde \psi_1(r,w)\  .
\end{equation}
Using Eqs. (\ref{Eq25}) and (\ref{Eq26}) it is easy to see that
\begin{equation}\label{Eq27}
W(we^{2 \pi i})=W(w)\  .
\end{equation}
Thus, using Eqs. (\ref{Eq25}),(\ref{Eq26}) and (\ref{Eq27}), we obtain the
relation
\begin{equation}\label{Eq28}
{{\tilde \psi_2(r,we^{2 \pi i})} \over {W(we^{2 \pi i})}} -
{{\tilde \psi_2(r,w)} \over {W(w)}} ={B \over A}\left(e^{4\pi i \beta} -1 
\right){{\Gamma(-2\beta-1)} \over {\Gamma(-\beta -2iw\alpha)}}
{{\tilde \psi_1(r,w)} \over {W(w)}}  .
\end{equation}
Since $W(w)$ is $r$-independent, we may use the large-$r$ asymptotic 
expansions of the confluent hypergeometric functions (given by Eqs. 13.5.1 and
13.5.2 in \cite{Abram}) in order to evaluate it. One finds
\begin{equation}\label{Eq29}
W(w)=-i {AB \Gamma (2 \beta +2) e^{ \pi i \beta} w^{-2 \beta -1} \over
{\Gamma (\beta +1-2iw \alpha) 2^{2 \beta +1}}}\  .
\end{equation}
(Of coarse, using the $|z| \to 0$ limit of the confluent hypergeometric 
functions, we obtain the same result). 
Thus, substituting (\ref{Eq28}) and (\ref{Eq29}) in (\ref{Eq24}) we obtain
\begin{equation}\label{Eq30}
G^C(y,x;t)={{i 2^{2\beta} \Gamma(-2\beta -1) \left(e^{3\pi i \beta}-
e^{- \pi i \beta} \right)} \over {\pi A^{2} \Gamma(2\beta+2)}} 
\int_{0}^{-i \infty} {{\Gamma(\beta+1-2iw \alpha)} \over {\Gamma(-\beta-
2iw \alpha)}} w^{2\beta+1} \tilde \psi_1(y,w) \tilde \psi_1(x,w) e^{-iwt} 
dw\  .
\end{equation}

\subsection{Asymptotic behaviour at timelike infinity}\label{Sec4B}

As was explained, the late-time behaviour of the field should follow from the
{\it low}-frequency contribution to the Green's function. Actually, it is easy 
to verify that the effective contribution to the integral in (\ref{Eq30}) 
should come from $|w|$=$O({1 \over t})$. Thus, in order to obtain the asymptotic
behaviour of the field at {\it timelike infinity} $i_+$ (where $x,y \ll t$),
we may use the $|w|r \ll 1$ limit of $\tilde \psi_1(r,w)$. Using Eq. 13.5.5
from \cite{Abram} one finds
\begin{equation}\label{Eq31}
\tilde \psi_1(r,w) \simeq Ar^{\beta +1}\  .
\end{equation}
Thus, we obtain
\begin{equation}\label{Eq32}
G^C(y,x;t)={{i2^{2 \beta} \Gamma (-2 \beta -1) \Gamma (\beta +1+ieQ)
(e^{3 \pi i \beta} - e^{-\pi i \beta})} \over {\pi \Gamma (2 \beta +2) 
\Gamma(- \beta +ieQ)}}(yx)^{\beta +1} \int_{0}^{-i \infty} w^{2 \beta +1}
e^{-iwt} dw\  ,
\end{equation}  
where we have used the relation $-2iw \alpha \simeq ieQ$ for $w \to 0$. 
Performing the integration in (\ref{Eq32}), one finds that
\begin{equation}\label{Eq33}
G^C(y,x;t)={{\Gamma(-2 \beta -1) \Gamma(\beta+1+ieQ)2^{2\beta +1} sin(2 \pi 
\beta)} \over {\pi \Gamma(- \beta+ieQ)}} x^{\beta +1} y^{\beta +1} t^{-(2 
\beta +2)}\  .
\end{equation}
Thus, the late-time behaviour of the 
{\it charged} scalar-field is dominated by the {\it electromagnetic} 
interaction (a {\it flat} spacetime effect) rather then by the spacetime 
curvature.
Moreover, we should point out an interesting and unique feature of this 
result. Contrary to neutral perturbations, where the amplitude of the late-time 
tail is {\it proportional} to the curvature of the spacetime (to $M$), 
the late-time behaviour of the {\it charged} scalar-field is {\it not} linear 
in the electromagnetic interaction (in the quantity $eQ$). In other words, the 
first Born approximation is {\it not} valid for general values of the 
quantity $eQ$ (in the first Born approximation, the amplitude to be 
backscattered and thus the late-time field itself, are {\it linear} in the 
scattering potential). The {\it physical} significance of this result is the
fact that the late-time behaviour of the {\it charged} scalar-field is 
dominated by {\it multiple} scattering from asymptotically far regions. This
is exactly the physical reason responsible for the fact that the analysis 
given in \cite{HodPir} is restricted to the $|eQ| \ll 1$ case (where the
first Born approximation is valid).  

\subsection{Asymptotic behaviour at future null infinity}\label{Sec4C}

Next, we go on to consider the behaviour of the charged scalar-field at
future null infinity $scri_+$ .
It is easy to verify that for this case
the effective frequencies contributing to integral (\ref{Eq30}) are
of order $O({1 \over u})$.
Thus, for $y-x \ll t \ll 2y-x$ one may use the
$|w|x \ll 1$ asymptotic limit for $\tilde \psi_1(x,w)$ and the $|w|y \gg 1$
($Imw < 0$) asymptotic limit of $\tilde \psi_1(y,w)$. Thus,
\begin{equation}\label{Eq34}
\tilde \psi_1(x,w) \simeq Ax^{\beta +1}\  ,
\end{equation}
and
\begin{equation}\label{Eq35}
\tilde \psi_1(y,w) \simeq Ae^{iwy} \Gamma (2 \beta +2) {{e^{-i {\pi \over 2}
(\beta +1-2iw \alpha)} (2w)^{-\beta-1+2iw \alpha} y^{2iw \alpha}} \over
{\Gamma (\beta+1+2iw \alpha)}}\  ,
\end{equation}
where we have used Eqs. 13.5.5 and 13.5.1 of \cite{Abram}, respectively.
Using this low-frequency limit one finds (for $v \gg u$)
\begin{equation}\label{Eq36}
G^C(y,x;t)={{\Gamma(-2 \beta -1) \Gamma(\beta+1+ieQ)2^{\beta} sin(2 \pi 
\beta)} \over {\pi \Gamma(- \beta+ieQ)}} x^{\beta +1} v^{-ieQ} u^{-( \beta
 +1-ieQ)}\  .
\end{equation}

\subsection{Asymptotic behaviour along the black-hole outer horizon}
\label{Sec4D}

Finally, we consider the behaviour of the charged
scalar-field at the black-hole outer-horizon $r_+$. While (\ref{Eq22})
and (\ref{Eq23}) are (approximated) solutions to the wave-equation
(\ref{Eq10}) in the $r \gg M,|Q|$ case, they do not represent the
solution near the horizon. As $y \to -\infty$ the wave-equation
(\ref{Eq10}) can be approximated by the equation
\begin{equation}\label{Eq37}
\tilde \psi _{,yy} + \left (w-{{eQ} \over {r_+}} \right ) ^2 \tilde \psi =0\  .
\end{equation}
Thus, we take
\begin{equation}\label{Eq38}
\tilde \psi_1(y,w)=C(w)e^{-i \left (w-{{eQ} \over {r_+}} \right )y}\  ,
\end{equation}
and we use (\ref{Eq34}) for $\tilde \psi_1(x,w)$. In order to match
the $y \ll -M$ solution with the $y \gg M$ solution we assume that the
two solutions have the same temporal dependence (this assumption has
been proven to be very successful for neutral perturbations
\cite{Gundlach}). In other words we take $C(w)$ to be $w$-independent.
In this case one should replace the roles of $x$ and $y$ in
Eqs. (\ref{Eq24}) and (\ref{Eq30}).
Using (\ref{Eq30}), we obtain
\begin{equation}\label{Eq39}
G^C(y,x;t)=\Gamma_{0} {{\Gamma(-2 \beta -1) \Gamma(\beta+1+ieQ)2^{2\beta +1} sin(2 \pi 
\beta)} \over {\pi \Gamma(- \beta+ieQ)}} e^{i {{eQ} \over {r_+}}y} 
v^{-(2 \beta +2)}\  ,
\end{equation}
where $\Gamma_{0}$ is a constant.

\subsection{The $|eQ| \ll 1$ case}\label{Sec4E}

Let us compare the results obtained in this paper with those obtained in 
\cite{HodPir}, using a {\it different} approach (for $|eQ| \ll 1$ ).
Taking the $|eQ| \ll 1$ limit of Eqs. (\ref{Eq33}),(\ref{Eq36}) and 
(\ref{Eq39}) one finds
\begin{equation}\label{Eq40}
G^C(y,x;t)={{2ieQ(-1)^{l}(2l)!!} \over {(2l+1)!!}} x^{\beta +1} y^{\beta +1}
t^{-(2 \beta +2)}\  ,
\end{equation}
at timelike infinity $i_+$,
\begin{equation}\label{Eq41}
G^C(y,x;t)={{ieQ(-1)^{l}l!} \over {(2l+1)!!}} x^{\beta +1} v^{-ieQ} 
u^{-(\beta +1-ieQ)}\  ,
\end{equation}
at future null infinity $scri_+$ and
\begin{equation}\label{Eq42}
G^C(y,x;t)=\Gamma_{0} {{2ieQ(-1)^{l}(2l)!!} \over {(2l+1)!!}} e^{i {{eQ} \over {r_+}}y}
v^{-(2 \beta +2)}\  ,
\end{equation}
along the black-hole outer-horizon $r_+$, respectively. 
These results have
exactly the same temporal and spatial dependence as those obtained in 
\cite{HodPir} for the $|eQ| \ll 1$ case (and for $v,t \ll |Q|e^{{1} \over
{|eQ|}}$).

\section{The Schwarzschild black-hole}\label{Sec5}

For a Schwarzschild black-hole ($Q=0, M \neq 0$) we have $\alpha =M$ and
$\beta =l$. Thus, in this case, expression (\ref{Eq30}) reduces to (taking
$M|w| \to 0 $)
\begin{equation}\label{Eq43}
G^C(y,x;t)={{4iM} \over {A^{2} \left [(2l+1)!! \right ]^2}} \int_{0}
^{-i \infty} w^{2l+2} \tilde \psi_1(y,w) \tilde \psi_1(x,w) e^{-iwt} dw\  .
\end{equation}
Now, taking the appropriate approximations for  $\tilde \psi_1(y,w)$ and 
$\tilde \psi_1(x,w)$ (as is done in Sec. \ref{Sec4}), one finds
\begin{equation}\label{Eq44}
G^C(y,x;t)={{(-1)^{l+1}4M(2l+2)!} \over {\left [(2l+1)!! \right ]^2}} x^{l+1}
y^{l+1} t^{-(2l+3)}\  ,
\end{equation}
at timelike infinity $i_+$,
\begin{equation}\label{Eq45}
G^C(y,x;t)={{(-1)^{l+1}2M(2l+2)!} \over {(2l+1)!!}} x^{l+1} u^{-(l+2)}\  ,
\end{equation}
at future null infinity $scri_+$, and
\begin{equation}\label{Eq46}
G^C(y,x;t)=\Gamma_{1} {{(-1)^{l+1}4M(2l+2)!} \over {\left [(2l+1)!! \right ]^2}} x^{l+1}
v^{-(2l+3)}\  ,
\end{equation}
along the black-hole outer horizon $r_+$, where $\Gamma_{1}$ is a constant.
Result (\ref{Eq44}) is just the one obtained by Leaver \cite{Leaver} and
subsequently by the simplified approach of \cite{Andersson}. Result 
(\ref{Eq45}) is the one obtained in \cite{Leaver}. Thus, the simplified 
approach of \cite{Andersson} can be {\it extended} to include also the 
asymptotic behaviour at future null infinity and along the future outer
horizon. Furthermore, result (\ref{Eq46}) is new in the sense that it
gives the $v^{-(2l+3)}$ dependence of the field along the black-hole outer
horizon using a {\it different} approach compared with the one given in 
\cite{Gundlach}.
Obviously, results (\ref{Eq44}),(\ref{Eq45}) and (\ref{Eq46}) are also valid
for the case of a {\it neutral} scalar-field, evolved on a 
Reissner-Nordstr\"om background, for then one also has $\alpha =M$ and
$\beta =l$.

\section{Summary and physical implications}\label{Sec6}
We have studied the asymptotic late-time behaviour of a {\it charged} 
gravitational collapse. Following the {\it no-hair theorem} 
we have focused attention
on the physical mechanism by which a {\it charged} hair is radiated away. 
The main results and their physical implications are:

\begin{enumerate}
\item Inverse {\it power-law} tails develop at timelike infinity, at null
infinity and along the black-hole outer horizon (where the power-law behaviour
is multiplied by an oscillatory term).

\item The dumping exponents for the {\it charged} field are {\it smaller}
compared with those of neutral perturbations. This implies that a black-hole
which is formed from a gravitational collapse of a charged matter becomes
``bald'' {\it slower} than a neutral one (due to the existence of charged
perturbations).

\item Since {\it charged} perturbations decay as a {\it power-law} they
are expected to cause a {\it mass-inflation} singularity during a charged
gravitational collapse which leads to the formation of a charged black-hole. 
Moreover, since charged perturbations decay {\it slower} than neutral ones
they are expected to {\it dominate} the mass-inflation
phenomena during a charged gravitational collapse. [This is caused by the
fact that the mass-function diverges like $m(v) \simeq v^{-p} e^{\kappa _0}$
, where ${1 \over 2} p$ is the dumping exponent of the field \cite{Poisson}.]

\item While the late-time behaviour of neutral perturbations is dominated by
the spacetime {\it curvature}, the late-time behaviour of {\it charged}
fields is dominated by {\it flat} spacetime effects, namely by the 
{\it electromagnetic interaction}.

\item While the amplitude of neutral tails is proportional to the the 
spacetime curvature (to $M$), the amplitude of {\it charged} tails is 
{\it not} linear in the electromagnetic interaction (in the quantity $eQ$).
This means that the late-time behaviour of {\it charged} perturbations is
physically determined by {\it multiple} scatterings (contrasted with neutral
perturbations).
Thus, a first Born approximation which is valid in the case of neutral
perturbations is {\it not} valid for {\it charged} ones. This is the physical
reason for which the analysis of paper I \cite{HodPir} is restricted to the
$|eQ| \ll 1$ case.

\item The results given in this paper can be reduced easily to the case of 
a Schwarzschild black-hole (or a neutral field on a Reissner-Nordstr\"om
background). For this case we demonstrate that one can obtain the asymptotic 
behaviour of the field along the black-hole outer horizon using the technique
of spectral decomposition.
\end{enumerate}

In accompanying forthcoming papers we study the {\it fully nonlinear} gravitational
collapse of a charged matter. We confirm {\it numerically} our {\it analytical}
predictions in the case of a {\it dynamically} charged gravitational
collapse which leads to a formation of a charged black-hole.

\bigskip
\noindent
{\bf ACKNOWLEDGMENTS}
\bigskip

This research was supported by a grant from the Israel Science Foundation.

\end{document}